\documentstyle[12pt]{article}
\begin{document}

\def\ds{\displaystyle}
\def\beq{\begin{equation}}
\def\eeq{\end{equation}}
\def\bea{\begin{eqnarray}}
\def\eea{\end{eqnarray}}
\def\beeq{\begin{eqnarray}}
\def\eeeq{\end{eqnarray}}
\def\ve{\vert}
\def\vel{\left|}
\def\ver{\right|}
\def\nnb{\nonumber}
\def\ga{\left(}
\def\dr{\right)}
\def\aga{\left\{}
\def\adr{\right\}}
\def\lla{\left<}
\def\rra{\right>}
\def\rar{\rightarrow}
\def\nnb{\nonumber}
\def\la{\langle}
\def\ra{\rangle}
\def\ba{\begin{array}}
\def\ea{\end{array}}
\def\tr{\mbox{Tr}}
\def\ssp{{\Sigma^{*+}}}
\def\sso{{\Sigma^{*0}}}
\def\ssm{{\Sigma^{*-}}}
\def\xis0{{\Xi^{*0}}}
\def\xism{{\Xi^{*-}}}
\def\qs{\la \bar s s \ra}
\def\qu{\la \bar u u \ra}
\def\qd{\la \bar d d \ra}
\def\qq{\la \bar q q \ra}
\def\gGgG{\la g^2 G^2 \ra}
\def\q{\gamma_5 \not\!q}
\def\x{\gamma_5 \not\!x}
\def\g5{\gamma_5}
\def\sb{S_Q^{cf}}
\def\sd{S_d^{be}}
\def\su{S_u^{ad}}
\def\ss{S_s^{??}}
\def\sbp{{S}_Q^{'cf}}
\def\sdp{{S}_d^{'be}}
\def\sup{{S}_u^{'ad}}
\def\ssp{{S}_s^{'??}}
\def\sig{\sigma_{\mu \nu} \gamma_5 p^\mu q^\nu}
\def\fo{f_0(\frac{s_0}{M^2})}
\def\ffi{f_1(\frac{s_0}{M^2})}
\def\fii{f_2(\frac{s_0}{M^2})}
\def\O{{\cal O}}
\def\sl{{\Sigma^0 \Lambda}}
\def\es{\!\!\! &=& \!\!\!}
\def\ar{&+& \!\!\!}
\def\ek{&-& \!\!\!}
\def\cp{&\times& \!\!\!}
\def\se{\!\!\! &\simeq& \!\!\!}
\def\kpm{&\pm& \!\!\!}
\def\kmp{&\mp& \!\!\!}
\def\arr{\!\!\!\!&+&\!\!\!}
% .........................................................

\def\simlt{\stackrel{<}{{}_\sim}}
\def\simgt{\stackrel{>}{{}_\sim}}

% .........................................................

\title{
         {\Large
                 {\bf
Fourth generation effects in processes induced by
$b \rar s$ transition 
                 }
         }
      }

\author{\vspace{1cm}\\
{\small T. M. Aliev$^a$ \thanks
{e-mail: taliev@metu.edu.tr}\,\,,
A. \"{O}zpineci$^b$ \thanks
{e-mail: ozpineci@ictp.trieste.it}\,\,,
M. Savc{\i}$^a$ \thanks
{e-mail: savci@metu.edu.tr}} \\
{\small a Physics Department, Middle East Technical University, 
06531 Ankara, Turkey}\\
{\small b  The Abdus Salam International Center for Theoretical Physics,
I-34100, Trieste, Italy} }
\date{}

\begin{titlepage}
\maketitle
\thispagestyle{empty}

\begin{abstract}
We study the effects of sequential fourth quark generation in rare
$B \rar K(K^\ast) \ell^+ \ell^-$ decays induced by $b \rar s$ transition 
and in $B_s^0 - \bar{B}_s^0$ mixing. Using the experimental values on branching
ratios of the $B \rar X_s \gamma$ and  $B \rar K(K^\ast) \ell^+ \ell^-$ decays, 
the allowed regions for $\vel V_{tb} V_{ts}^\ast\ver$ and 
$\vel V_{t^\prime b} V_{t^\prime s}^\ast\ver$ are determined as a function
of the $t^\prime$ quark mass.
\end{abstract}

%\vspace{1cm}
~~~PACS numbers: 12.60.--i, 13.20.--v, 13.20.He
\end{titlepage}

\section{Introduction}

Despite Standard Model (SM) describes successfully all low energy experiments,
it is an incomplete theory. This theory contains many unsolved and fundamental
problems, such as the origin of CP violation, mass spectrum and the number
of generations. Recent observation of neutrino oscillations \cite{R5301}
indicated that neutrino sector of SM must be enlarged. One of the most
straightforward and economical extension of the SM is adding the fourth
generation to the fermionic sector, similar to the three--generation case.
The extra generation can contribute to the electroweak radiative
corrections. Latest studies in the electroweak sector allow the existence of
a fourth generation with heavy Dirac neutrino \cite{R5302,R5303}. Moreover, two or
three extra generations with relatively "light" neutrinos, with a mass about
$50~GeV$ are also allowed \cite{R5303}. Flavor--changing neutral current
(FCNC) transitions provide potentially the most sensitive and stringiest test
for the SM at loop level, since they are forbidden in the SM at tree
level. At the same time these transitions are very sensitive to the new
physics beyond the SM via contributions of the new particles to the loop
diagrams. It should be stressed that if newly proposed particles are heavy and 
if they cannot be produced directly in the accelerators, their influence
through the loop diagrams can be a unique possibility for establishing new
physics beyond the SM. The effects of the fourth generation to rare decays
have been studied in many works \cite{R5304}--\cite{R5308}.

Although theoretically FCNC processes are highly suppressed in the SM, very 
exciting results are obtained on the experimental side. The first measurement 
of the FCNC processes through $b \rar s \gamma$ were reported by CLEO \cite{R5309}.
Recently, $B \rar K \ell^+ \ell^-$ decay is observed at B factories at SLAC 
and KEK \cite{R5310}--\cite{R5312}. BaBar Collaboration also reported their
preliminary results about the observation of $B \rar K^\ast \ell^+ \ell^-$
decay with branching ratio ${\cal B}(B \rar K^\ast \ell^+ \ell^-) =
\ga 1.68^{+0.68}_{-0.58} \pm 0.28 \dr \times 10^{-6}$ and $90\%$ C.L. 
${\cal B}(B \rar K^\ast \ell^+ \ell^-) <3 \times 10^{-6}$ \cite{R5312}. In
this paper, we study the contributions of the fourth generation to the
processes induced by the $b \rar s$ transitions and use the experimental
results of the branching ratio for the $b \rar s \gamma$ \cite{R5313}, 
$B \rar K \ell^+ \ell^-$ and $B \rar K^\ast \ell^+ \ell^-$ decays and try to
determine the constraints on the extended Cabibbo--Kobayashi--Maskawa matrix
(CKM) elements $\vel V_{tb} V_{ts}^\ast\ver$ and $\vel V_{t^\prime b} 
V_{t^\prime s}^\ast\ver$.

The paper is organized as follows. In section 2, we present the basic theoretical 
expressions for the differential widths of the $B \rar K \ell^+ \ell^-$ and 
$B \rar K^\ast \ell^+ \ell^-$ decays, and for the  mass difference 
$\Delta m_{B_s}$, with sequential up--like quark in the fourth generation model. 
Section 3 is devoted to the numerical analysis and the conclusion.

\section{Theoretical results}

In this section we present the necessary theoretical formulae for the 
$B \rar X_s \gamma$, $B \rar K \ell^+ \ell^-$, $B \rar K^\ast \ell^+ \ell^-$
decays and for the mass difference in the $B_s^0 - \bar{B}_s^0$ system in 
presence of the fourth generation. All these processes, except mixing in the
$B_s^0 - \bar{B}_s^0$ system, are induced by the $b \rar s$ transition. 
At quark level this transition is described by the effective Hamiltonian
\bea
\label{e1}
{\cal H}_{eff} = \frac{\alpha G_F}{2 \sqrt{2} \pi} V_{tb} V_{ts}^\ast
\sum_{i=1}^{10} C_i(\mu) {\cal O}_i(\mu)~,
\eea
where the full set of operators in ${\cal O}_i(\mu)$ and the corresponding
expressions for the Wilson coefficients in the SM3 (here and in all further 
discussions SM3 and SM4 will denote SM with three and four generations, 
respectively) are given in \cite{R5314,R5315}. As is well known, the fourth
generation introduces the first three generations into SM copiously, and
hence it is clear that it changes only values of the Wilson coefficients
$C_7(\mu)$, $C_9(\mu)$ and $C_{10}(\mu)$ with the help of the running fourth 
generation up quark $t^\prime$ at loop level, and do not introduce any new 
operator structure, i.e.,
\bea
\label{e2}
C_7^{tot}(\mu) &=& C_7^{SM}(\mu) + \frac{V_{t^\prime b}V_{t^\prime s}^\ast}
{V_{tb} V_{ts}^\ast} C_7^{t^\prime} (\mu) ~, \nnb \\
C_9^{tot}(\mu) &=& C_9^{SM}(\mu) + \frac{V_{t^\prime b}V_{t^\prime s}^\ast}
{V_{tb}V_{ts}^\ast} C_9^{t^\prime} (\mu) ~, \nnb \\
C_{10}^{tot}(\mu) &=& C_{10}^{SM}(\mu) + \frac{V_{t^\prime b}
V_{t^\prime s}^\ast}
{V_{tb}V_{ts}^\ast} C_{10}^{t^\prime} (\mu) ~,
\eea
where $V_{t^\prime b}$ and $V_{t^\prime s}$ are the elements of the $4\times 4$
Cabibbo--Kobayashi--Maskawa (CKM) matrix. The explicit forms of the
$C_i^{t^\prime}$ can easily be obtained from the
SM results by simply substituting $m_t \rar m_{t^\prime}$. 
Neglecting the $s$ quark mass, the effective Hamiltonian leads to the
following matrix element for the $b \rar s \ell^+ \ell^-$ transition
\bea
\label{e3}
{\cal M} &=& \frac{G\alpha}{2\sqrt{2} \pi}
 V_{tb}V_{ts}^\ast
\Bigg[ C_9^{tot} \, \bar s \gamma_\mu (1-\gamma_5) b \,
\bar \ell \gamma_\mu \ell +
C_{10}^{tot} \bar s \gamma_\mu (1-\gamma_5) b \,
\bar \ell \gamma_\mu \gamma_5 \ell \nnb \\
&-& 2  C_7^{tot}\frac{m_b}{q^2} \bar s \sigma_{\mu\nu} q^\nu
(1+\gamma_5) b \, \bar \ell \gamma_\mu \ell \Bigg]~,
\eea
where $q^2=(p_1+p_2)^2$ and $p_1$ and $p_2$ are the four--momenta of the final
leptons. We observe from Eq. (\ref{e3}) that in order to calculate
the matrix element for the $B \rar K^\ast(K)\ell^+ \ell^-$ decay, the matrix
elements of the quark operators in Eq. (\ref{e3}) need to be sandwiched
between initial and final $(K$ or $K^\ast)$ meson states, which results in a
form that is parametrized in terms of the form factors    
\bea
\lefteqn{
\label{e4}
\lla K^\ast(p_{K^\ast},\varepsilon) \vel \bar s \gamma_\mu
(1 - \gamma_5) b \ver B(p_B) \rra =} \nnb \\
&& - i \varepsilon_\mu^\ast (m_B+m_{K^\ast})A_1(q^2)
+ i (p_B + p_{K^\ast})_\mu (\varepsilon^\ast q)
\frac{A_2(q^2)}{m_B+m_{K^\ast}} \nnb \\
&&+i q_\mu \frac{2 m_{K^\ast}}{q^2} (\varepsilon^\ast q)
\left[A_3(q^2)-A_0(q^2)\right] 
- \epsilon_{\mu\nu\lambda\sigma} \varepsilon^{\ast\nu} p_{K^\ast}^\lambda
q^\sigma
\frac{2 V(q^2)}{m_B+m_{K^\ast}}~,
\eea
where $\varepsilon$ is the
polarization vector of $K^\ast$ meson and $q = p_B-p_{K^\ast}$ is the momentum 
transfer. Using the equation of motion, the form factor $A_3(q^2)$ can be
written in terms of $A_1(q^2)$ and $A_2(q^2)$ as follows
\bea
\label{e5}
A_3(q^2) = \frac{(m_B+m_{K^\ast})}{2 m_{K^\ast}} A_1(q^2) -
\frac{(m_B-m_{K^\ast})}{2 m_{K^\ast}} A_2(q^2)~.
\eea
In order to ensure that there exists no kinematical singularity we assume
that $A_3(q^2=0) = A_0(q^2=0)$. 

The corresponding form factors are defined 
through  the matrix elements for the $B \rar K$ transition as
\bea
\label{e6}
\lla K(p_{K}) \vel \bar s \gamma_\mu b \ver B(p_B) \rra =
f_+ \Bigg[ (p_B+p_K)_\mu - \frac{m_B^2-m_K^2}{q^2} \, q_\mu \Bigg]
+ f_0 \,\frac{m_B^2-m_K^2}{q^2} \, q_\mu~.
\eea
Finetness of Eq. (\ref{e6}) is guaranteed by demanding $f_+(0) = f_0(0)$.

The semileptonic form factors for the $K^\ast$ and $K$ mesons resulting from
the dipole operator $\bar s i \sigma_{\mu\nu} q^\nu (1 + \gamma_5) b$ are
defined as
\bea
\lefteqn{
\label{e7}
\lla K^\ast(p_{K^\ast},\varepsilon) \vel \bar s i \sigma_{\mu\nu} q^\nu
(1 + \gamma_5) b \ver B(p_B) \rra =} \nnb \\
&&2 \epsilon_{\mu\nu\lambda\sigma}
\varepsilon^{\ast\nu} p_{K^\ast}^\lambda q^\sigma
T_1(q^2) + i \left[ \varepsilon_\mu^\ast (m_B^2-m_{K^\ast}^2) -
(p_B + p_{K^\ast})_\mu (\varepsilon^\ast q) \right] T_2(q^2) \nnb \\
&&+ i (\varepsilon^\ast q) \left[ q_\mu -
(p_B + p_{K^\ast})_\mu \frac{q^2}{m_B^2-m_{K^\ast}^2} \right]
T_3(q^2)~,
\eea
\bea
\label{e8}
\lla K(p_{K}) \vel \bar s i\sigma_{\mu\nu} q^\nu(1 + \gamma_5)
 b \ver B(p_B) \rra = - \, \frac{f_T}{m_B+m_K}
\Big[ (p_B+p_K)_\mu q^2 -
q_\mu (m_B^2-m_K^2)\Big]~.
\eea
The matrix elements of the $B \rar K \ell^+ \ell^-$ and
$B \rar K^\ast \ell^+ \ell^-$ decays can be written as

\bea
\label{e9}
{\cal M} = \frac{G \alpha}{2 \sqrt{2} \pi} V_{tb} V_{ts}^\ast
m_B \ga F_\mu^{1i} \bar \ell \gamma_\mu \ell +    
F_\mu^{2i} \bar \ell \gamma_\mu \gamma_5 \ell \dr~,
\eea
where $i=1$ corresponds to $K$ meson and $i=2$ corresponds to $K^\ast$
meson, respectively, and
\bea
\label{e10}
F_\mu^{11} \es \frac{1}{m_B} \Big[A^\prime (p_B+p_K)_\mu + B^\prime q_\mu \Big]~,\\
\label{e11}
F_\mu^{21} \es \frac{1}{m_B} \Big[C^\prime (p_B+p_K)_\mu  + D^\prime q_\mu \Big]~, \\
\label{e12}
F_\mu^{12} \es \Big[
- \frac{1}{m_B^2} A \epsilon_{\mu\nu\rho\sigma} \varepsilon^{\ast\nu}
p_{K^\ast}^\rho q^\sigma
 -i B_1 \varepsilon_\mu^\ast
+ i \frac{1}{m_B^2} B_2 (\varepsilon^\ast q) p_{K^\ast\mu}
+ i \frac{1}{m_B^2} B_3 (\varepsilon^\ast q) q_\mu  \Big]~, \\
\label{e13}
F_\mu^{22} \es \Big[
- \frac{1}{m_B^2} C_1 \epsilon_{\mu\nu\rho\sigma} \varepsilon^{\ast\nu}
p_{K^\ast}^\rho q^\sigma
 -i D_1 \varepsilon_\mu^\ast
+ i \frac{1}{m_B^2} D_2 (\varepsilon^\ast q) p_{K^\ast\mu}    
+ i \frac{1}{m_B^2} D_3 (\varepsilon^\ast q) q_\mu  \Big]~,     
\eea
where
\bea
\label{e14}
A^\prime \es C_9^{tot} f_+ + 2 \frac{m_b}{m_B}\frac{1}{(1+\sqrt{r})} 
C_7^{tot} f_T~, \nnb \\
B^\prime \es C_9^{tot} f_+ + 2 \frac{m_b}{m_B}\frac{(1-\sqrt{r})}{s} 
C_7^{tot} f_T~, \nnb \\
C^\prime \es C_{10}^{tot} f_+~, \nnb \\
D^\prime \es C_{10}^{tot} f_-~, \nnb \\
A \es \frac{2 V}{1+\sqrt{r}} C_9^{tot} + 4\frac{m_b}{m_B s} 
C_7^{tot} T_1~, \nnb \\
B_1 \es (1+\sqrt{r})\Big[C_9^{tot} A_1 + 2\frac{m_b}{m_B s} (1-\sqrt{r}) 
C_7^{tot} T_2\Big]~, \nnb \\
B_2 \es \frac{1}{1-r}\Big[(1-\sqrt{r}) C_9^{tot} A_2 + 2\frac{m_b}{m_B}
C_7^{tot} \Big( T_3 + \frac{1-r}{s} T_2 \Big)\Big] ~, \nnb \\
B_3 \es \frac{1}{s}\Big[2 \sqrt{r} C_9^{eff} (A_3-A_0) -
2\frac{m_b}{m_B} C_7^{tot} T_3 \Big]~, \nnb \\
C_1 \es \frac{2 V}{1+\sqrt{r}} C_{10}^{tot}~, \nnb \\
D_1 \es (1+\sqrt{r}) C_{10}^{tot} A_1~, \nnb \\
D_2 \es \frac{A_2}{1+\sqrt{r}} C_{10}^{tot}~, \nnb \\
D_3 \es \frac{2\sqrt{r}}{s} (A_3-A_0)  C_{10}^{tot}~,  
\eea
where 
\bea
f_-(q^2) = \frac{(1-r)}{s}\Big[ f_0(q^2)-f_+(q^2) \Big]~,~~ 
r = \frac{m_{K^\ast}^2}{m_B^2}~,~~s = \frac{q^2}{m_B^2}~.\nnb
\eea

Using the matrix element, for the dilepton invariant mass
distribution we get
\bea
\label{e15}
\frac{d\Gamma^{B\rar K}}{d s} \es \frac{G^2 \alpha^2 m_B^5}{2^{10} \pi^5}
\vel V_{tb} V_{ts}^\ast \ver^2 \sqrt{\lambda} \, v
\Bigg\{\frac{\lambda}{3} (3-v^2) \ga \vel A^\prime \ver^2 + \vel C^\prime \ver^2 \dr
+ s (1-v^2) (2 + 2 r-s) \vel C^\prime \ver^2 \nnb \\
\ar 2 s (1-v^2) (1-r) \mbox{\rm Re}[C^\prime D^{\prime\ast}] + 
s^2 (1-v^2) \vel D^\prime \ver^2 \Bigg\}~,
\eea
\bea                                                     
\label{e16}
\frac{d\Gamma^{B\rar K^\ast}}{d s} \es \frac{G^2 \alpha^2 m_B^5}{2^{10} \pi^5}
\vel V_{tb} V_{ts}^\ast \ver^2 \sqrt{\lambda} \, v
\Bigg\{\frac{1}{6} s \lambda \ga 3-v^2 \dr \vel A \ver^2
+ \frac{1}{3} s \lambda v^2 \vel C_1 \ver^2 \nnb \\ 
\ar \frac{1}{12 r} \Bigg[ (3-v^2) \ga \lambda + 12 r s \dr  \vel B_1 \ver^2
+\ga \lambda (3-v^2) + 24 r s v^2 \dr  \vel D_1 \ver^2 \Bigg]\nnb \\
\ar \frac{\lambda}{12 r} \Bigg[ \lambda (3-v^2) \vel B_2 \ver^2
+\Big\{ \lambda (3-v^2) + 3 (1-v^2) s (2+2 r-s)\Big\}  \vel D_2 \ver^2 \Big)\nnb \\
\ek \frac{\lambda}{6r} \Bigg[(3-v^2) (1-r-s) \mbox{\rm Re}
[B_1 B_2^\ast] + \Big\{ (3-v^2) (1-r-s) + 3 (1-v^2) s
\Big\} \mbox{\rm Re}[D_1 D_2^\ast] \Bigg] \nnb \\
\ek \frac{\lambda}{2 r}(1-v^2) s \Big( \mbox{\rm Re}[D_1 D_3^\ast] - (1-r)
\mbox{\rm Re}[D_2 D_3^\ast] \Big) + \frac{\lambda}{4 r} (1-v^2) s^2 \vel D_3
\ver^2 \Bigg\}~,
\eea
where $v^2 = 1 - 4m_\ell^2/q^2$ and $\lambda(a,b,c) = a^2+b^2+c^2-2ab-2ac-2bc$ 
is the usual triangle function.

In constraining up quark type fourth generation effects, we will also
consider
$B^0_s-\bar{B}^0_s$ mixing. The mass difference $\Delta m_{B_s}$ in SM4 can
be written as
\bea
\label{e17}
\Delta m_{B_s} \es \frac{G^2 m_W^2}{6\pi^2} m_{B_s} B_{B_s} f_{B_s}^2 \Big\{
\eta_t \ga V_{tb} V_{ts}^\ast \dr^2 S_0(x_t) + 
\eta_{t^\prime} \ga V_{t^\prime b} V_{t^\prime s}^\ast \dr^2
S_0(x_{t^\prime}) \nnb \\
\ar 2 \eta_{t t\prime} \ga V_{tb} V_{ts}^\ast \dr \ga V_{t^\prime b} V_{t^\prime s}^\ast
\dr S(x_t,x_{t^\prime}) \Big\}~,
\eea
where $x_t=m_t^2/m_W^2$, $x_{t^\prime}=m_{t^\prime}^2/m_W^2$ and
\bea
\label{e18}
S_0(x_t) \es \frac{4 x_t - 11 x_t^2 + x_t^3}{4 (1-x_t)^2}
-\frac{3}{2} \frac{x_t^3 \mbox{\rm ln} x_t}{(1-x_t)^3}~,\\
\label{e19}
S_0(x_{t^\prime}) \es S_0(x_t \rar x_{t^\prime})~, \\
\label{e20}
S(x_t,x_{t^\prime}) \es x_t x_{t^\prime} \Bigg\{
\frac{1}{x_{t^\prime}-x_t} \Bigg[
\frac{1}{4} + \frac{3}{2} \frac{1}{1-x_{t^\prime}} -
\frac{3}{4} \frac{1}{(1-x_{t^\prime})^2} \Bigg]  
\mbox{\rm ln} x_{t^\prime} \nnb \\
\ek \frac{1}{x_{t^\prime}-x_t} \Bigg[
\frac{1}{4} + \frac{3}{2} \frac{1}{1-x_t} -
\frac{3}{4} \frac{1}{(1-x_t)^2} \Bigg]  
\mbox{\rm ln} x_t \nnb \\
\ek \frac{3}{4} \frac{1}{(1-x_t) (1-x_{t^\prime})} \Bigg\}~.
\eea
Here $\eta_t=0.55$ is the QCD correction factor. Taking into account the
threshold effect from $b^\prime$ quark, $\eta_{t t^\prime}$ is estimated to be
\cite{R5307}
\bea
\eta_{t t^\prime} = \Big(\alpha_s(m_t)\Big)^{6/23} 
\ga \frac{\alpha_s(m_{b^\prime})}{\alpha_s(m_t)} \dr^{6/21} 
\ga \frac{\alpha_s(m_{t^\prime})}{\alpha_s(m_{b^\prime})} \dr^{6/19}~.\nnb
\eea
Note that when $m_{t^\prime}$ lies between $250~GeV$ and $400~GeV$, 
$\eta_{t t^\prime}$ is
quite close to $\eta_{t^\prime}$ numerically, hence for simplicity, in further 
analysis we will set $\eta_{t t^\prime}=\eta_{t^\prime}$.  

In order to obtain quantitative results the value of the fourth generation
CKM matrix element $V_{t^\prime b} V_{t^\prime s}^\ast$ is needed. For
this aim we will use the experimentally measured values of the branching
ratios ${\cal B}(B \rar X_s \gamma)$ and ${\cal B}(B \rar X_c e
\bar{\nu}_e)$. To eliminate the uncertainty coming from $b$ quark
mass we consider the ratio
\bea
\label{e21}
R = \frac{{\cal B}(B \rar X_s \gamma)}{{\cal B}(B \rar X_c e \bar{\nu}_e)}~.
\eea
In leading logarithmic approximation this ratio is equal to
\bea
\label{e22}
R = \frac{6 \alpha \vel C_7^{tot}(m_b)V_{tb} V_{ts}^\ast \ver^2}
{\pi f(\hat m_c) \kappa(\hat m_c) \vel V_{cb} \ver^2} ~,
\eea
where $\hat m_c = m_c/m_b$ and the functions $f(\hat m_c)$ and
$\kappa(\hat m_c)$ for the $b \rar c \ell \bar{\nu}$ transition
are given by \cite{R5316} 
\bea
\label{e23}
f(\hat m_c) \es 1 - 8 \hat m_c^2 + 8 \hat m_c^6 - \hat m_c^8 -
24 \hat m_c^4 \,\mbox{\rm ln} (\hat m_c) ~, \nnb \\ \nnb \\
\kappa(\hat m_c) \es 1 - \frac{2 \alpha_s(m_b)}{3 \pi}
\Bigg[ \ga \pi^2 - \frac{31}{4} \dr \ga 1 - \hat m_c^2 \dr^2 + \frac{3}{2}
\Bigg]~.
\eea
From Eqs. (\ref{e21}) and (\ref{e22})  we get
\bea
\label{e24}
\vel C_7^{SM} V_{tb} V_{ts}^\ast +
C_7^{new} V_{t^\prime b} V_{t^\prime s}^\ast \ver =
\sqrt{
\frac{\pi f(\hat m_c) \kappa(\hat m_c) \vel V_{cb} \ver^2}{6 \alpha}~
\frac{{\cal B}(B \rar X_s \gamma)}{{\cal B}(B \rar X_c e \bar \nu_e)}}~.
\eea
The model parameters can be constrained from the measured branching ratios 
of the $B \rar X_s \gamma$ decay and ${\cal B}(B \rar X_c e \bar{\nu}_e)= 
10.4\%$
\bea
{\cal B}(B \rar X_s \gamma) = \left\{ \begin{array}{lc}
\left( 3.21 \pm 0.43 \pm 0.27^{+0.18}_{-0.10} \right) \times
10^{-4}& \cite{R5317}~,\\ \\   
\left( 3.36 \pm 0.53 \pm 0.42 \pm 0.54 \right) \times
10^{-4}& \cite{R5318}~,\\ \\   
\left( 3.11 \pm 0.80 \pm 0.72\right) \times
10^{-4}& \cite{R5319}~.\end{array} \right. \nnb
\eea

In our numerical analysis, we will use the weighted average value
${\cal B}(B \rar X_s \gamma) = (3.23 \pm 0.42)\times 10^{-4}$ \cite{R5320}
for the branching ratio of the $B \rar X_s \gamma$ decay.

Another constraint to the extended CKM matrix element comes from the
unitarity condition, i.e., 
\bea
\label{e25}
\vel V_{us} \ver^2 + \vel V_{cs} \ver^2 + \vel V_{ts} \ver^2 +
\vel V_{t^\prime s} \ver^2 \es 1~,\nnb \\
\vel V_{ub} \ver^2 + \vel V_{cb} \ver^2 + \vel V_{tb} \ver^2 +
\vel V_{t^\prime b} \ver^2 \es 1~,\nnb \\
V_{ub} V_{us}^\ast + V_{cb} V_{cs}^\ast + V_{tb} V_{ts}^\ast +
V_{t^\prime b} V_{t^\prime s}^\ast \es 0 ~.
\eea

Since charged--current tree--level decays are well measured experimentally 
they are not affected by new physics at leading order. Therefore, 
for the parameters $\vel V_{us} \ver$, $\vel V_{cs} \ver$, $\vel V_{cb}
\ver$ and $\vel V_{ub}/V_{cb} \ver$ we will make use of    
Particle Data Group (PDG) constraints \cite{R5313}.
Using the weighted average for ${\cal B} (B \rar X_s \gamma)$ and PDG 
constraint $0.38 \le \vel V_{cb} \ver \le 0.044$,
from Eqs. (\ref{e24}) and (\ref{e25}) we get the
constraints
\bea
\label{e26}
0.011 \le \vel C_7^{SM} V_{tb} V_{ts}^\ast \right. \arr \left. C_7^{new}
V_{t^\prime b}
V_{t^\prime s}^\ast \ver \le 0.015~,\\
\label{e27}
0.03753 \le \vel V_{tb} V_{ts}^\ast \right. \arr \left. V_{t^\prime b}
V_{t^\prime s}^\ast
\ver \le 0.043976~,\\
\label{e28}
0 \le \vel V_{ts} \ver^2 \arr \vel V_{t^\prime s} \ver^2 \le 0.00492~,\\
\label{e29}
0.998 \le \vel V_{tb} \ver^2 \arr  \vel V_{t^\prime b} \ver^2 \le 0.9985~.
\eea

\section{Numerical analysis}

In this section we will study the constraints to 
$\vel V_{t^\prime b} V_{t^\prime s}^\ast \ver$ coming from the measured
branching ratios of the $B \rar K \ell^+ \ell^-$ and $B \rar K^\ast
\ell^+ \ell^-$ decays and $B_s^0-\bar{B}_s^0$ mixing, as well as using
the results in Eqs. (\ref{e26})--(\ref{e29}). The main input parameters
involved in calculation of the branching ratios of the $B \rar K \ell^+
\ell^-$ and $B \rar K^\ast \ell^+ \ell^-$ decays are the form factors, whose 
values we take from light cone QCD sum rule \cite{R5321}--\cite{R5323},
where the form factors are expressed in terms of three parameters as       
\bea
F(s) = \frac{F(0)}{\ds 1-a_F s + b_F s^2}~, \nnb
\eea
where the values of parameters $F(0)$, $a_F$ and $b_F$ for the
$B \rar K$ and $B \rar K^\ast$ decay are listed in Table 1.

\begin{table}[h]
\renewcommand{\arraystretch}{1.5}
\addtolength{\arraycolsep}{3pt}
$$
\begin{array}{|l|ccc|}
\hline
& F(0) & a_F & b_F \\ \hline
f_+^{B \rar K} &
0.35 & 1.37 & 0.35 \\
f_0^{B \rar K} &
0.35 & 0.40 & 0.41 \\
A_1^{B \rar K^*} &
0.337 & 0.60 & -0.023 \\
A_2^{B \rar K^*} &
0.283 & 1.18 & \phantom{-}0.281\\
A_0^{B \rar K^*} &
0.470 & 1.55 & \phantom{-}0.68\phantom{0}\\
V^{B \rar K^*}   &
0.458 & 1.55 & \phantom{-}0.575\\
T_1^{B \rar K^*} &
0.379 & 1.59 & \phantom{-}0.615\\
T_2^{B \rar K^*} &
0.379 & 0.49 & -0.241\\
T_3^{B \rar K^*} &
0.261 & 1.20 & \phantom{-}0.098\\ \hline           
\end{array}        
$$       
\caption{$B$ meson decay form factors in a three-parameter fit, where the
radiative corrections to the leading twist contribution and SU(3) breaking
effects are taken into account (see \cite{R5322,R5323}).}
\renewcommand{\arraystretch}{1}
\addtolength{\arraycolsep}{-3pt}
\end{table}

The values of the other input parameters which we use in our numerical
calculations are: $m_b=4.8~GeV$, $m_c=1.35~GeV$, $m_{B_s}=5.369~GeV$,
$\tau_{B_s}=1.64\times 10^{-12}~s$ and $B_{B_s} f_{B_s}^2 = (0.26~GeV)^2$.
The experimental lower bound of the mass difference is $\Delta m_{B_s} \ge
14.9~ps^{-1}$. For the values of the Wilson coefficients $C_7^{SM}$, 
$C_9^{SM}$ and $C_{10}^{SM}$ we have used their next--to--leading
logarithmic result: $C_7^{SM}=-0.308$, $C_9^{SM}=4.154$ and
$C_{10}^{SM}=-4.261$. It should be noted that the decays $B \rar K \ell^+
\ell^-$ and $B \rar K^\ast \ell^+ \ell^-$ receive long distance contribution
coming from $\bar c c$ intermediate states. In the
present work we neglect such long distance effects. The strong dependence
on $m_{t^\prime}$ (see for example \cite{R5304}) makes the electroweak
penguins a good place looking for the existence of fourth generation.
Contributions of fourth generation to $B \rar K(K^\ast) \ell^+ \ell^-$ decays   
have already been studied (see second references in \cite{R5304} and
\cite{R5306}). The present investigation differs from the above--mentioned
works in two aspects: 
\begin{itemize}
\item we use the experimentally measured results on branching ratio,
\item we consider $V_{tb}V_{ts}^\ast$ and 
$V_{t^\prime b}V_{t^\prime s}^\ast$ as two independent complex parameters,
which were taken to be real in \cite{R5304} and
\cite{R5306}.
\end{itemize}

The complex parameters $V_{tb}V_{ts}^\ast$ and 
$V_{t^\prime b}V_{t^\prime s}^\ast$ are constrained by the unitarity conditions
(see Eqs. (\ref{e27})--(\ref{e29})), the measured branching ratios $B \rar X_s
\gamma$ (see Eq. (\ref{e26})) and $B \rar K(K^\ast) \ell^+ \ell^-$ decays
(see Eqs. (\ref{e15}), (\ref{e16})), which depend on $m_{t^\prime}$. For each value of
$m_{t^\prime}$ there exists an allowed region in the $\vel
V_{tb}V_{ts}^\ast \ver$--$\vel V_{t^\prime b}V_{t^\prime s}^\ast \ver$  
plane. 

Since there exists no analytical solution of Eqs. (\ref{e26})--(\ref{e29}), 
we will solve these
equations numerically assuming that $V_{tb}V_{ts}^\ast$ and
$V_{t^\prime b}V_{t^\prime s}^\ast$ are complex. For sufficiently large
number of randomly chosen complex parameters $V_{t^\prime b}V_{t^\prime
s}^\ast$ and $V_{tb}V_{ts}^\ast$, the selected values would range over the whole 
solution space. In Figs. (1)--(3) we present the allowed region for
$V_{tb}V_{ts}^\ast$ and $V_{t^\prime b}V_{t^\prime s}^\ast$ at
$m_{t^\prime}=200~GeV$, $m_{t^\prime}=300~GeV$ and $m_{t^\prime}=400~GeV$,
respectively. In obtaining this solution region we have used Eqs.
(\ref{e26})--(\ref{e29}) and Eq. (\ref{e15}). From these figures we see
that, for $\vel V_{t^\prime b}V_{t^\prime s}^\ast \ver=0$,
$\vel V_{tb}V_{ts}^\ast\ver$ takes on values that is close to the SM
prediction and is mainly distributed around $\sim 0.04$.
Moreover, when $\vel V_{t^\prime b}V_{t^\prime s}^\ast \ver$ increases, the
allowed region of $\vel V_{tb}V_{ts}^\ast\ver$ becomes wider with the center
being fixed around $0.04$, and the values of $\vel V_{t^\prime b}V_{t^\prime
s}^\ast \ver$ are mainly distributed around $0.01$. With increasing values
of $m_{t^\prime}$, the allowed region for 
$\vel V_{t^\prime b}V_{t^\prime s}^\ast \ver$ becomes narrower. This
behavior can be explained as follows. Wilson coefficients are strongly
dependent on $m_{t^\prime}$ and in order to remain in the experimentally allowed
region, the element $\vel V_{t^\prime b}V_{t^\prime s}^\ast \ver$ of CKM
matrix must decrease, since branching ratio contains factors like 
$\vel C_i V_{tb}V_{ts}^\ast\ver$. Similar behavior is observed for the 
$B \rar K^\ast \ell^+ \ell^-$ decay (see Figs. (4)--(6)).

In Figs. (7) and (8) we present the dependence of $\Delta m_{B_s}$ on 
$\vel V_{tb}V_{ts}^\ast\ver$ and $\vel V_{t^\prime b}V_{t^\prime s}^\ast
\ver$, taking into account the lower experimental bound for
$\Delta m_{B_s}$, at two different values of $m_{t^\prime}$.
It follows from both figures that the main distribution is in the range
$0.36 \le \vel V_{tb}V_{ts}^\ast\ver \le 0.044$ and 
$0\le\vel V_{t^\prime b}V_{t^\prime s}^\ast \ver\le 0.01$. With increasing
values of $m_{t^\prime}$, obviously, $\Delta m_{B_s}$ also increases.

Finally we would like to note that restrictions to 
$\vel V_{t^\prime d}V_{t^\prime b}^\ast \ver$ matrix element can be obtained
by an investigation of the rare decays induced through $b \rar d$
transition. Further analysis of the decays induced by $b \rar s(d)$
transition is more promising in studying new sources for CP violation, since 
$4\times 4$ CKM matrix predicts the existence of three CP violating phases.
We will discuss this issue elsewhere in future. 

In conclusion, we have studied the effect of the fourth generation quark to
the rare decays induced by FCNC $b \rar s$ transition. Using the
experimental result for the branching ratios of the $B \rar X_s \gamma$, 
$B \rar K(K^\ast) \ell^+ \ell^-$ decays and the unitarity condition for the
$4\times 4$ CKM matrix, we have determined the allowed parameter space for 
 $\vel V_{tb}V_{ts}^\ast\ver$ and $\vel V_{t^\prime b}V_{t^\prime s}^\ast
\ver$ in their dependence on $m_{t^\prime}$.

\newpage

\section*{Figure captions}
{\bf Fig. (1)} Three--dimensional plot of the branching ratio for the
$B \rar K \ell^+ \ell^-$ decay, with
respect to the allowed parameter space of $V_{tb}V_{ts}^\ast$ and
$V_{t^\prime b}V_{t^\prime s}^\ast$, at $m_{t^\prime}=200~GeV$.\\ \\ 
{\bf Fig. (2)} The same as in Fig. (1), but at $m_{t^\prime}=300~GeV$.\\ \\ 
{\bf Fig. (3)} The same as in Fig. (1), but at $m_{t^\prime}=400~GeV$.\\ \\
{\bf Fig. (4)} The same as in Fig. (1), but for the            
$B \rar K^\ast \ell^+ \ell^-$ decay.\\ \\
{\bf Fig. (5)} The same as in Fig. (4), but at $m_{t^\prime}=300~GeV$\\ \\
{\bf Fig. (6)} The same as in Fig. (4), but at $m_{t^\prime}=400~GeV$.\\ \\
{\bf Fig. (7)} Three--dimensional plot of the mass 
difference $\Delta m_{B_s}$ of the $B_s^0-\bar{B}_s^0$ system, with
respect to the allowed parameter space of $V_{tb}V_{ts}^\ast$ and
$V_{t^\prime b}V_{t^\prime s}^\ast$, at $m_{t^\prime}=300~GeV$.\\ \\
{\bf Fig. (8)} The same as in Fig. (7), but at $m_{t^\prime}=700~GeV$.\\ \\

\newpage

\begin{figure}
\vskip 1.5 cm
    \includegraphics{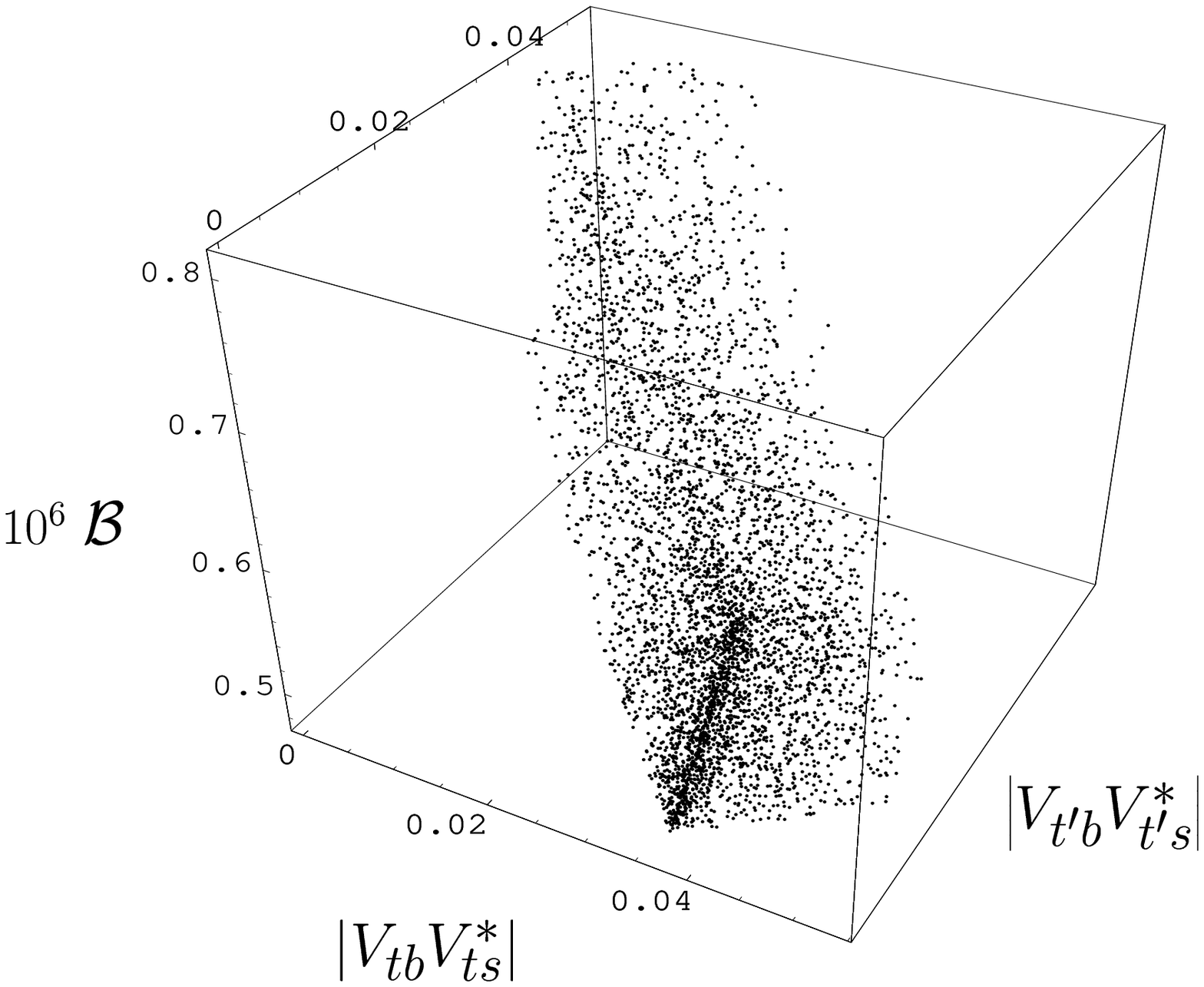}
\vskip 13.25cm
\caption{}
\vskip 10cm
%\begin{center}
%{\bf Fig. 1--a} 
%\end{center}
\end{figure}  

\begin{figure}   
\vskip 2.5 cm
    \includegraphics{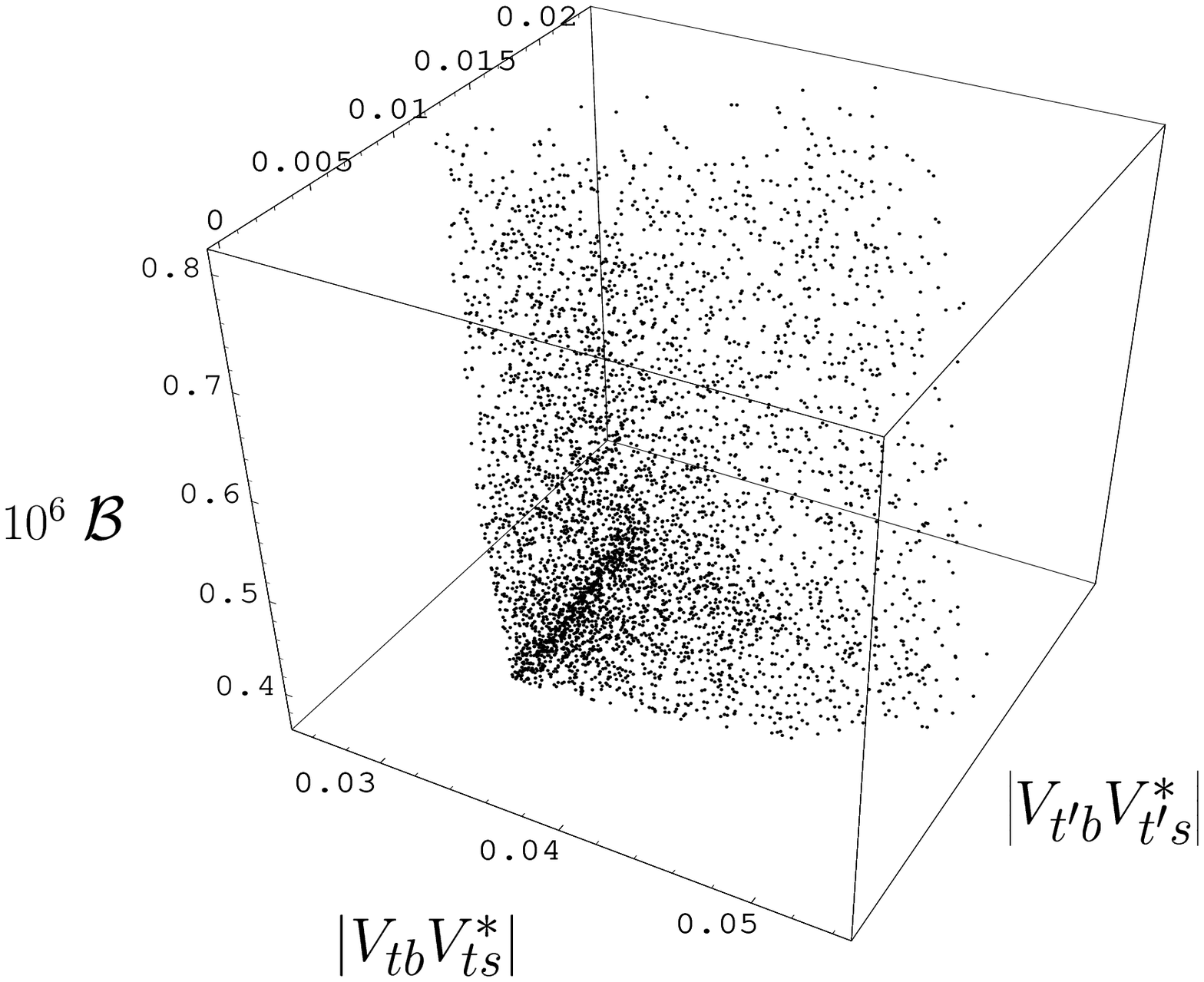}
\vskip 13.25 cm   
\caption{}
\vskip 10cm
%\begin{center}
%{\bf Fig. 1--b}
%\end{center}
\end{figure}

\begin{figure}   
\vskip 1.5 cm
    \includegraphics{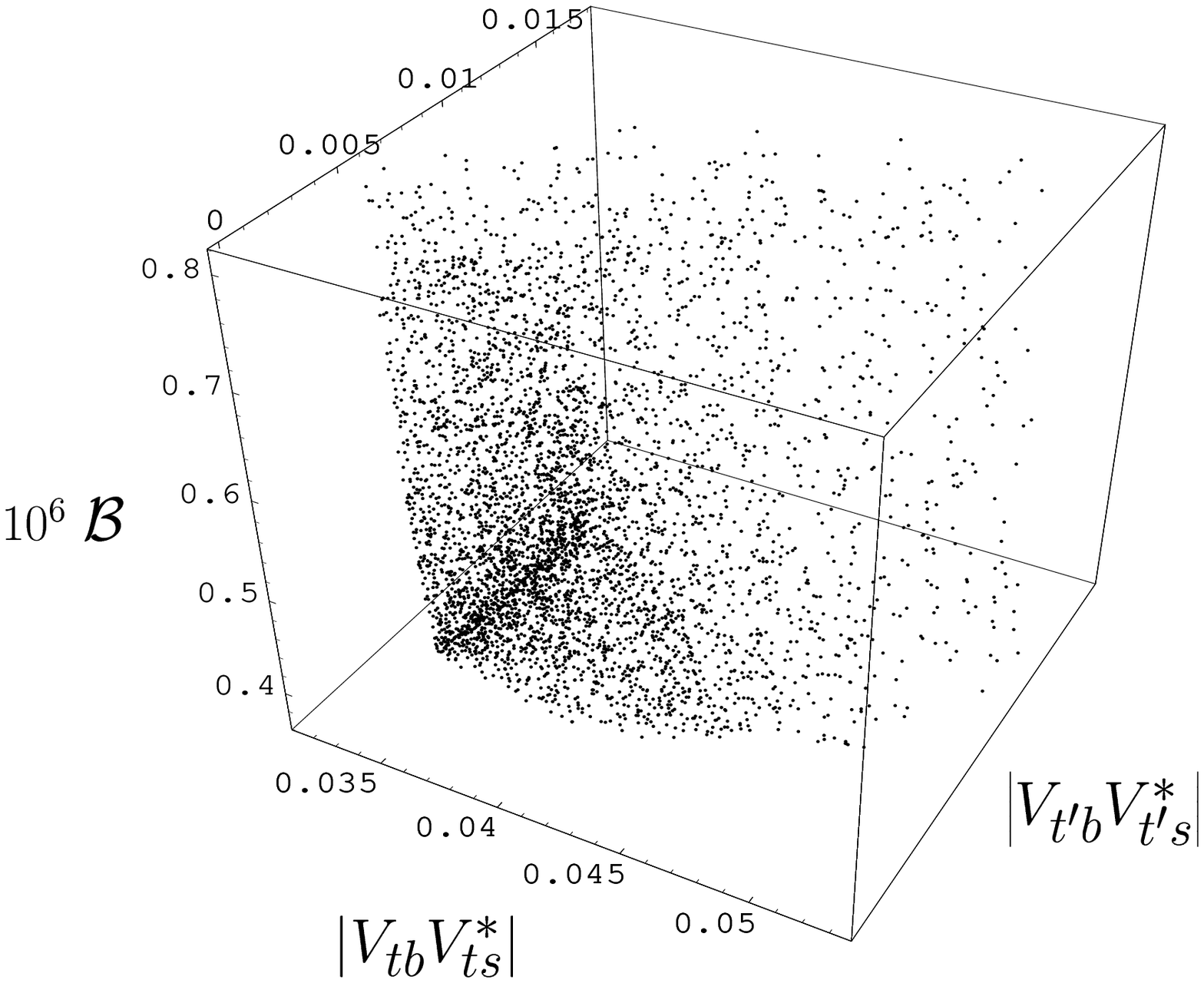}
\vskip 13.25cm
\caption{}
\vskip 10cm
%\begin{center}
%{\bf Fig. 2--a}
%\end{center}
\end{figure}

\begin{figure}    
\vskip 2.5 cm
    \includegraphics{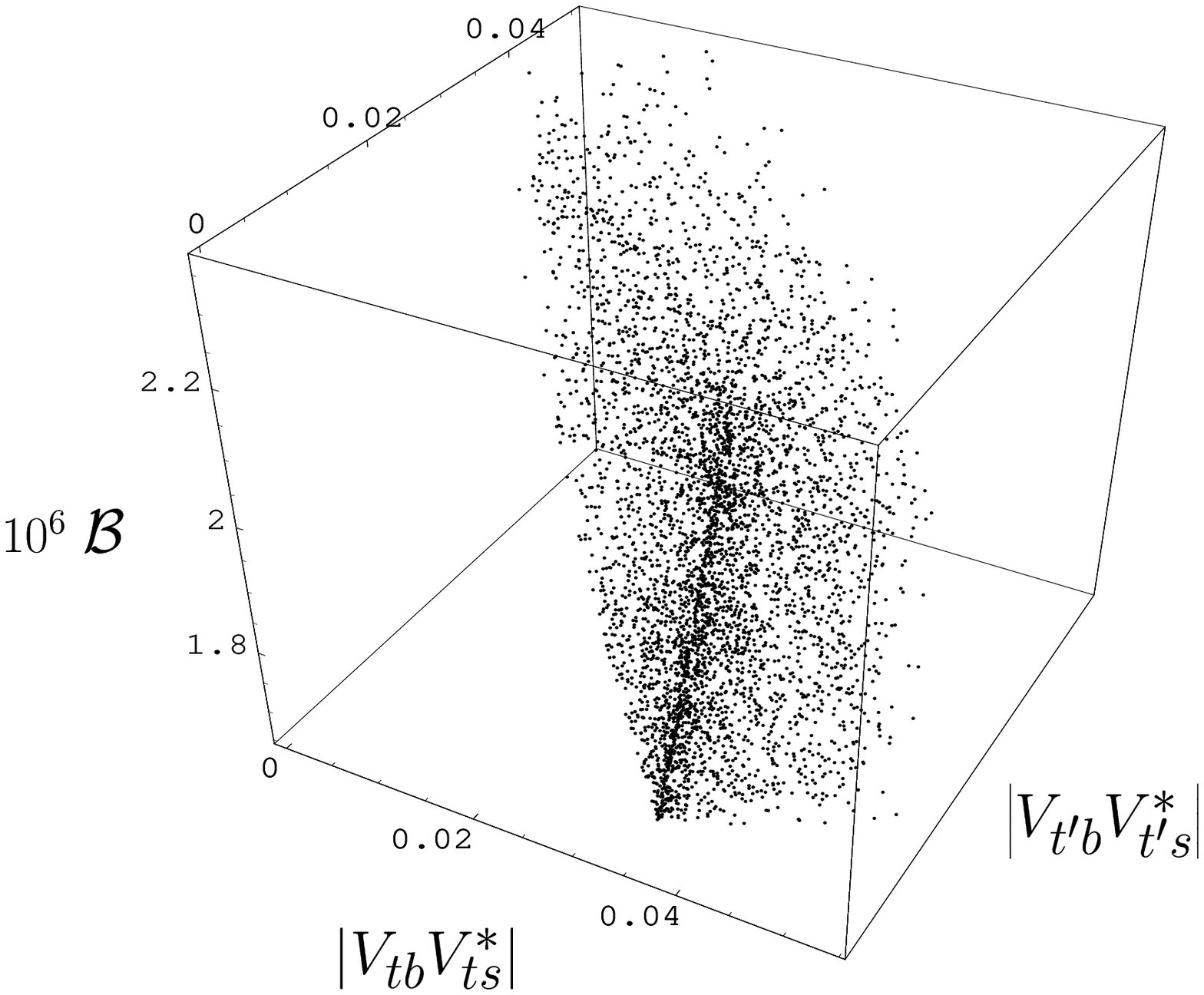}
\vskip 13.25 cm   
\caption{}
\vskip 10cm
%\begin{center}
%{\bf Fig. 2--b}
%\end{center}
\end{figure}

\begin{figure}
\vskip 1.5 cm
    \includegraphics{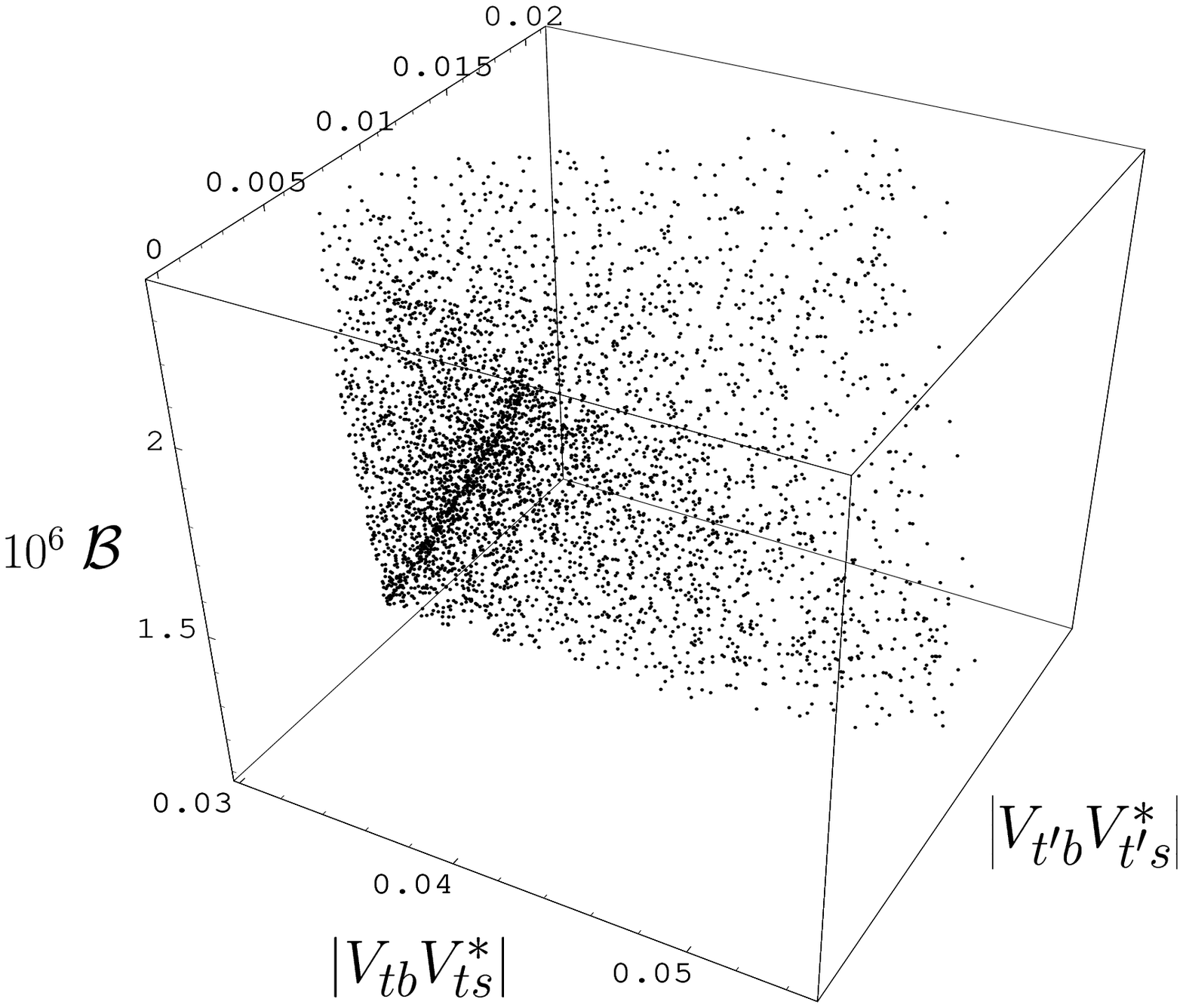}
\vskip 13.25cm
\caption{}
\vskip 10cm
%\begin{center}
%{\bf Fig. 1--a} 
%\end{center}
\end{figure}  

\begin{figure}   
\vskip 2.5 cm
    \includegraphics{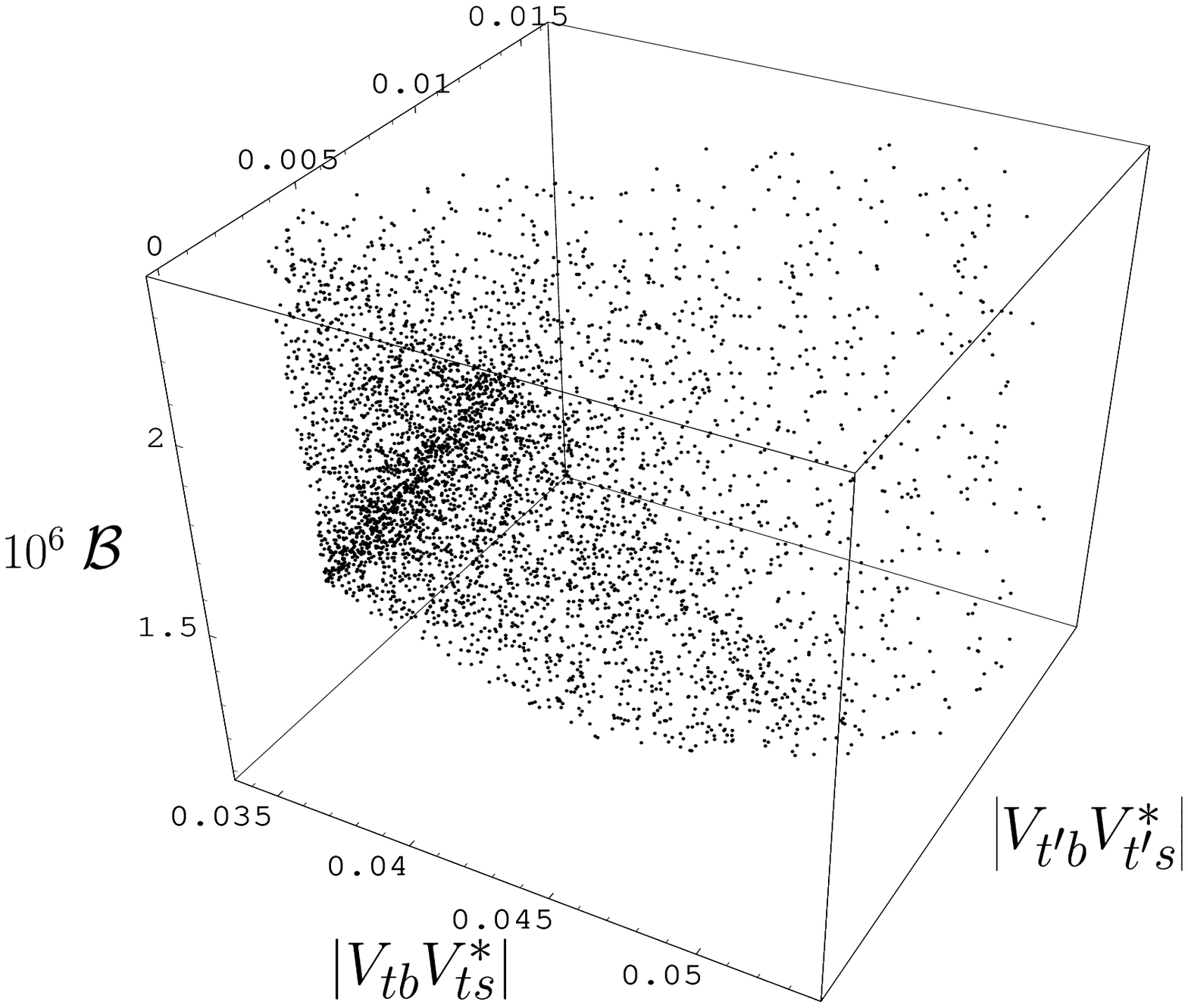}
\vskip 13.25 cm   
\caption{}
\vskip 10cm
%\begin{center}
%{\bf Fig. 1--b}
%\end{center}
\end{figure}

\begin{figure}   
\vskip 1.5 cm
    \includegraphics{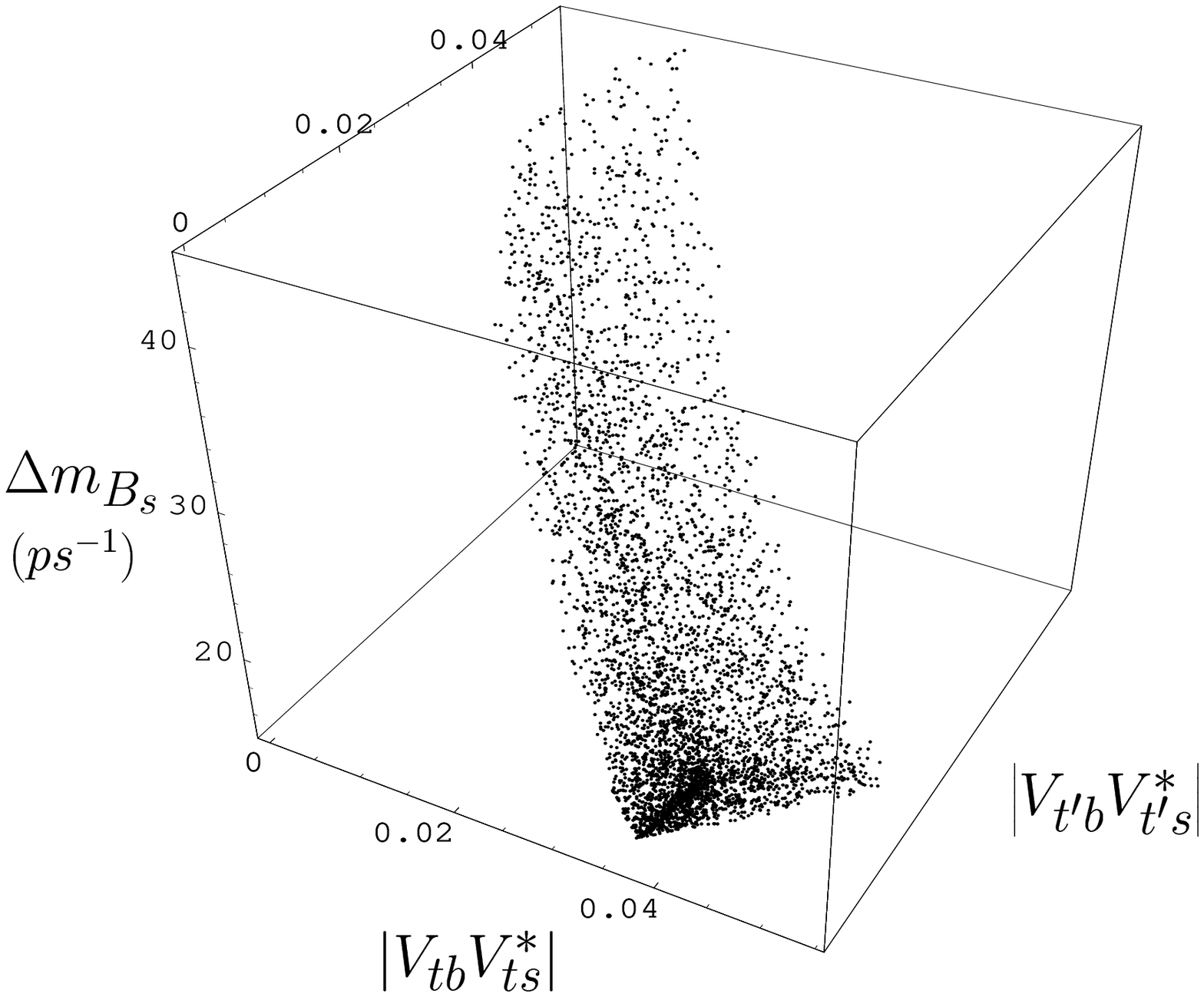}
\vskip 13.25cm
\caption{}
\vskip 10cm
%\begin{center}
%{\bf Fig. 2--a}
%\end{center}
\end{figure}

\begin{figure}    
\vskip 2.5 cm
    \includegraphics{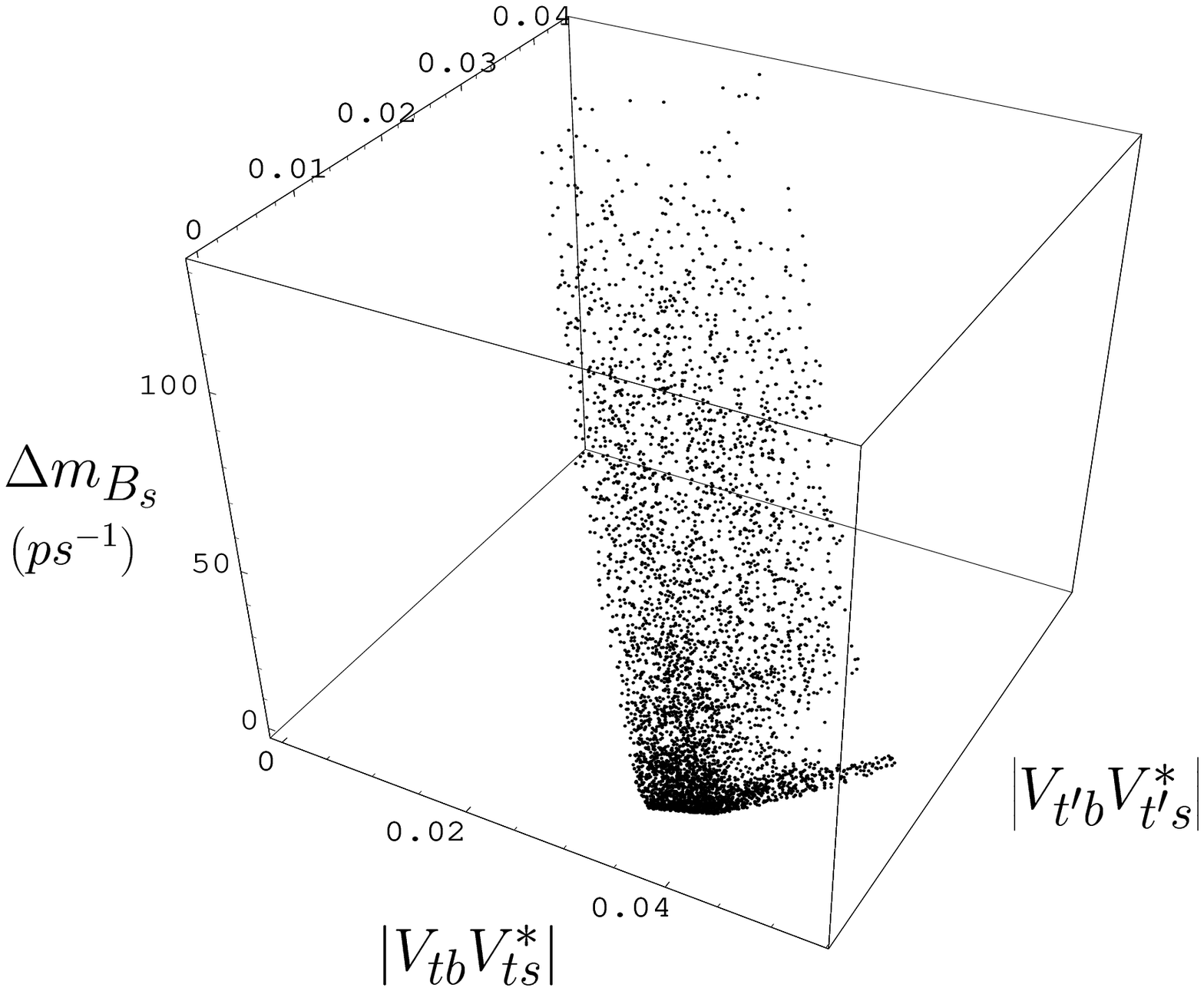}
\vskip 13.25 cm   
\caption{}
\vskip 10cm
%\begin{center}
%{\bf Fig. 2--b}
%\end{center}
\end{figure}

\end{document}